\documentclass[preprint,
superscriptaddress,
amsmath,amssymb,aps,showkeys,showpacs,
twoside,final,secnumarabic,
nofootinbib]{revtex4-2}

\usepackage[paperwidth=205mm,paperheight=290mm,top=17mm,bottom=25mm,
inner=17mm,outer=17mm,
twoside]{geometry}

\usepackage{cmap} 
\usepackage[T1,T2A]{fontenc}
\usepackage[utf8]{inputenc}
\usepackage[russian,english]{babel}
\usepackage{color}
\usepackage{graphicx}
\usepackage{dcolumn}
\usepackage{bm} 
\usepackage[unicode=true,colorlinks=true,linkcolor=magenta, urlcolor=blue, citecolor = blue,breaklinks]{hyperref}
\usepackage{multirow}
\usepackage{url}
\usepackage{breakurl}
\usepackage{subcaption}

\newlength{\dhatheight}
\newcommand{\doublehat}[1]{
    \settoheight{\dhatheight}{\ensuremath{\hat{#1}}}
    \addtolength{\dhatheight}{-0.35ex}
    \hat{\vphantom{\rule{1pt}{\dhatheight}}
    \smash{\hat{#1}}}
}

\newcommand{\br}[1]{\left(#1\right)}

\newcommand{\cbr}[1]{\left\{#1\right\}} 
\newcommand{\prob}[2]{P\br{#1\,\middle|\,#2}}

\newcommand{\bhh}{\doublehat{b}}
\DeclareGraphicsExtensions{.eps}

\newcount\issue
\newcount\Vol
\newcount\numb
\usepackage{fancyhdr} 
\def\Vol{\textbf{80}}
\def\numb{x}
\begin{document}

\title{StatTestCalculator: A New General Tool \\ for Statistical Analysis \\ in High Energy Physics} 

\def\addressa{Moscow State University, Faculty of Physics}
\def\addressb{Skobeltsyn Institute of Nuclear Physics, M.V.~Lomonosov Moscow State University}

\author{\firstname{E.}~\surname{Abasov}}
\affiliation{\addressa}
\affiliation{\addressb}
\author{\firstname{L.V.}~\surname{Dudko}}
\affiliation{\addressb}
\author{\firstname{D.E.}~\surname{Gorin}}
\affiliation{\addressa}
\affiliation{\addressb}
\author{\firstname{O.S.}~\surname{Vasilevskii}}
\affiliation{\addressa}
\affiliation{\addressb}

\begin{abstract}
We present \textit{StatTestCalculator} (STC), a new open-source statistical analysis tool designed for analysis high energy physics experiments. STC provides both asymptotic calculations and Monte Carlo simulations for computing the exact statistical significance of a discovery or for setting upper limits on signal model parameters. We review the underlying statistical formalism, including profile likelihood ratio test statistics for discovery  and exclusion hypotheses, and the asymptotic distributions that allow quick significance estimates. We explain the relevant formulas for the likelihood functions, test statistic distributions, and significance metrics $Z$ (both with and without incorporating systematic uncertainties). The implementation and capabilities of STC are described, and we validate its performance against the widely-used CMS \textsc{Combine} tool. We find excellent agreement in both the expected discovery significances and upper limit calculations. STC is a flexible framework that can accommodate  systematic uncertainties and user-defined statistical models, making it suitable for a broad range of analyses.
\end{abstract}

\pacs{Suggested PACS}\par
\keywords{Suggested keywords   \\[5pt]}

\maketitle
\thispagestyle{fancy}


\section{Introduction}\label{intro}

Experiments in high energy physics (HEP) rely on statistical analysis of observed data to draw conclusions about new phenomena. Because the outcomes of collider experiments are inherently probabilistic, rigorous statistical methods are required to estimate parameters and assess the significance of potential discoveries. Many sophisticated statistical tools have been developed for HEP analyses, such as the \textsc{RooFit} and \textsc{RooStats} frameworks~\cite{roo_fit},\cite{roo_stats}, the CMS \textsc{Combine} tool~\cite{combine}, \textsc{Theta}~[4], HistFactory \cite{histfactory} and others. These tools are powerful but often designed for complex, large-scale analyses. In contrast, there is a need for a tool that enables quick yet accurate general-purpose statistical calculations for a variety of common scenarios.

In this work, we introduce \textit{StatTestCalculator} (STC), a new lightweight software package for statistical analysis in HEP. STC is specifically developed on Python and is designed to be easy to use and versatile. STC allows to run statystical analysis on data presented as distributions, such as binned histograms obtained from an experiment or neural network's (NN) discriminator outputs. Since STC is developed on Python it can be easily incorporated in NN pipeline to automatically run an analysis of the results. The key capabilities of STC are:
\begin{itemize}
    \item Calculation of expected discovery significance and upper limits using asymptotic formulae (closed-form approximations).
    \item Calculation of exact significance values and upper limits through Monte Carlo simulations of test statistic distributions.
    \item Incorporation of systematic uncertainties for both signal and background yields, with options for normal, lognormal or user-defined distributions of systematic effects.
    \item Support for specifying  systematic uncertainties that are presented as a shape-of-distribution uncertainty or a number of events in a single bin uncertainty.
    \item Extensibility for user-defined test statistics, formulas, or custom systematic uncertainty distributions.
    \item Tools for visualizing data and fitted signal/background distributions, as well as plotting the distributions of test statistics from Monte Carlo simulations.
\end{itemize}
Since STC is written in pure Python, one can easily understand and modify the code to suit specific needs. In the following sections, we outline the mathematical foundations of the statistical tests implemented in STC, describe the implementation and usage of the software, demonstrate and test its performance on example problems. We particularly focus on the consistency of STC's results with established methods (such as \textsc{Combine}) for both significance estimation and confidence limit setting.

\section{Mathematical Background}

\subsection{\label{sec:level2}Statistical Hypotheses and Likelihood Formalism}
In the context of collider experiments, we consider two hypotheses when searching for a new signal atop a background process. The \textbf{null hypothesis} $H_0$ (background-only) assumes that the data contain no contribution from the new signal, whereas the \textbf{alternative hypothesis} $H_1$ (signal+background) assumes that signal events are present in addition to background. We define a signal strength parameter $\mu$ such that $\mu=0$ corresponds to $H_0$ and $\mu=1$ corresponds to the nominal signal prediction under $H_1$. The goal of a discovery test is to reject $H_0$ when data are more compatible with $H_1$. Conversely, to set an upper limit (exclusion test), one may attempt to reject a specific non-zero $\mu$ (often $\mu=1$ representing the expected signal) in favor of $\mu=0$ if the data show no significant excess.

To formalize the problem, we construct a likelihood function $L(\mu,\theta)$ for the observed data, where $\theta$ denotes any nuisance parameters (e.g. expected background rate or other systematic parameters) that are not of direct interest but must be accounted for. For example, in a simple counting experiment with N signal regions, the observed count $n_i$ can be assumed Poisson-distributed: $n_i \sim \mathrm{Poisson}(\mu s_i + b_i)$, where $s_i$ is the expected number of signal events (for $\mu=1$) and $b_i$ is the expected background yield (which might be uncertain). If $b_i$ is known exactly, the likelihood is 
\[
L(\mu) = \prod_{i=1}^N\frac{(\mu s_i + b_i)^{n_i} e^{-(\mu s_i + b_i)}}{n_i!}\,, 
\] 

In more general scenarios, $\theta$ may include $b$ and other parameters. Consider the statistical model of the data obtained:
\begin{equation}
L(\mu, \theta) = \prod_{i=1}^N\frac{(\mu s_i + \tau_i b_i)^{n_i} e^{-(\mu s_i + b_i)}}{n_i!} \prod_{j=1}^M Systematic(\theta)
\label{eq:likelihood}
\end{equation}
Here Systematic is the distribution of systematic uncertainty and $\theta$ represents the vector of nuisance parameters. $\tau$ is distributed according to a systematic distributions:

\begin{equation}
\tau \sim Systematic = \frac{1}{\sqrt{2\pi}\delta}e^{-\frac{(x-1)^2}{2\delta^2}} - \text{normal distribution}
\end{equation}
or
\begin{equation}
\tau \sim Systematic = \frac{e^{-\frac{[\ln(x)-1]^2}{\delta^2}}}{x\delta\sqrt{2\pi}}- \text{lognormal distribution}
\end{equation}

\subsection{Profile Likelihood Ratios and Test Statistics}
To test a given hypothesis on $\mu$, we use the profile likelihood ratio as our test statistic. The \textbf{profile likelihood ratio} is defined as 
\begin{equation}
\lambda(\mu) = \frac{L(\mu, \hat{\hat{\theta}}(\mu))}{L(\hat{\mu}, \hat{\theta})}\,,
\label{eq:plr}
\end{equation}
where $\hat{\mu}$ and $\hat{\theta}$ are the values of $\mu$ and $\theta$ that maximize the likelihood (the unconditional maximum likelihood estimates, MLEs), and $\hat{\hat{\theta}}(\mu)$ denotes the value of $\theta$ that maximizes $L$ for a fixed $\mu$. In other words, the numerator is the maximum likelihood achievable when $\mu$ is fixed to the hypothesis of interest, and the denominator is the absolute maximum likelihood across all parameters. By construction, $0 \le \lambda(\mu) \le 1$, with larger values indicating that the hypothesis $\mu$ is more compatible with the data.

It is often more convenient to work with the logarithm of $\lambda$. We define the test statistic $q_\mu$ as 
\begin{equation}
q_\mu = -2 \ln \lambda(\mu) = -2 \ln \frac{L(\mu, \hat{\hat{\theta}}(\mu))}{L(\hat{\mu}, \hat{\theta})}\,.
\label{eq:qmu_def}
\end{equation}
Large values of $q_\mu$ imply that the data are in tension with the hypothesis $\mu$, since the likelihood at $\mu$ is significantly lower than the maximum likelihood. We will use two special cases of $q_\mu$:
\begin{itemize}
    \item \textbf{Discovery test statistic $q_0$:} This is $q_\mu$ with $\mu=0$. It is used to test $H_0$ (no signal) against the alternative of some $\mu>0$. We define $q_0$ with an additional one-sided convention~\cite{cowan}:
    \begin{equation}
    q_0 = 
    \begin{cases}
    -2\ln \lambda(0)~, & \hat{\mu} \ge 0,\\
    0~, & \hat{\mu} < 0~,
    \end{cases}
    \label{eq:q0_def}
    \end{equation}
    such that $q_0$ is set to zero if the best-fit $\hat{\mu}$ is negative (unphysical for a positive signal strength). In practice, a negative $\hat{\mu}$ would indicate a downward fluctuation in the data or a mismodeling, rather than evidence of new signal, so taking $q_0=0$ in that case ensures the test statistic does not falsely imply a significant discovery for deficits.
    \item \textbf{Exclusion test statistic $q_\mu$ (for upper limits):} In the context of setting an upper limit on the signal strength, one typically tests a specific $\mu > 0$ (e.g. $\mu=1$ for the nominal signal) against the background-only alternative. We consider 
    \begin{equation}
    q_\mu = 
    \begin{cases}
    -2\ln \lambda(\mu)~, & \hat{\mu} \le \mu,\\
    0~, & \hat{\mu} > \mu~,
    \end{cases}
    \label{eq:qmu_def2}
    \end{equation}
    so that $q_\mu$ is only nonzero when the fitted signal strength $\hat{\mu}$ does not exceed the tested value. If the data favor a signal even larger than $\mu$ (i.e. $\hat{\mu}>\mu$), then that data outcome is not considered evidence against the hypothesis $\mu$ (one would not exclude $\mu$ if the data trend high, since a higher $\mu$ would fit even better). Thus $q_\mu$ is set to 0 in that case to reflect that $\mu$ is not excluded by an upward fluctuation.
\end{itemize}

\subsection{Discovery Significance: Analytical Formula}
For discovery, we are interested in the median significance with which we can reject the background-only hypothesis $H_0$ if a true signal of strength $\mu'=1$ exists. Starting from the likelihood in Eq.~(\ref{eq:likelihood}) with one signal region and one control region (incorporating a background uncertainty via $b$ and $m$), we derive the formula for the test statistic $q_0$ on the Asimov dataset. First, we find the maximum likelihood estimates by differentiating $\ln L(\mu,b)$ and setting derivatives to zero. Solving these, one finds for the simple one-bin signal + control case~\cite{gorin}:
\begin{align}
\hat{\mu} &= \frac{n - m/\tau}{s}\,, \label{eq:muhat}\\
\hat{b} &= \frac{m}{\tau}\,, \label{eq:bhat}\\
\hat{\hat{b}}(\mu) &= \frac{n + m - (1-\tau)\mu s}{2(1+\tau)}\,, \label{eq:bhathat}
\end{align}
where $\hat{\hat{b}}(\mu)$ is the conditional MLE of $b$ assuming a fixed signal strength $\mu$. Using these in the profile likelihood ratio for $\mu=0$, we obtain $q_0 = -2\ln \lambda(0)$. After some algebraic simplification, the expression for $q_0$ in terms of the observed counts $n$ and $m$ is:
\begin{equation}
q_0 = 
\begin{cases}
-2 \Big[ n \ln \frac{n+m}{(1+\tau)n} + m \ln \frac{\tau(n+m)}{(1+\tau)m} \Big]\,, & n \ge \frac{m}{\tau},\\[2ex]
0\,, & n < \frac{m}{\tau}~,
\end{cases}
\label{eq:q0_nm}
\end{equation}
which is consistent with Eq.~(\ref{eq:q0_def}) that $q_0$ vanishes if $\hat{\mu}<0$ (the condition $n < m/\tau$ corresponds to a negative $\hat{\mu}$ in Eq.~\ref{eq:muhat}). 

To find the \emph{expected} discovery significance, we now evaluate $q_0$ on the Asimov dataset for $\mu'=1$. This means we substitute $n = s + b$ and $m = \tau b$ (the expected counts when signal is present at strength 1). In that case, $\hat{\mu}=1$ and $n \ge m/\tau$ is automatically satisfied, so we use the first line of Eq.~(\ref{eq:q0_nm}). The Asimov value of the test statistic, $q_{0,A}$, then yields the median discovery test statistic. The median discovery significance (often denoted $Z_{\text{disc}}$) is:
\begin{equation}
Z_{\text{disc}} = \sqrt{q_{0,A}} = \sqrt{-2\Big[ (s+b)\ln\frac{(s+b)(1+\tau b)}{(1+\tau)(s+b)\,b} + \tau b \ln\left(1 + \frac{s}{(1+\tau) b}\right) \Big]}\,.
\end{equation}
We can simplify this expression. Note that $\tau b$ is the expected value of $m$, and it is related to the variance of $\hat{b}$. In fact, one can define a fractional background uncertainty $\delta$ by 
\begin{equation}
\delta = \frac{\sigma_b}{b} = \frac{1}{\sqrt{\tau b}}\,, 
\end{equation}
since $\mathrm{Var}(\hat{b}) = b/\tau$ for the control measurement, so $\sigma_b = \sqrt{b/\tau}$. Rewriting the above expression in terms of $\delta$, we obtain a more interpretable formula for the discovery significance including the effect of a background systematic uncertainty:
\begin{equation}
Z_{\text{disc}} = \sqrt{\,2 \Bigg[ (s+b)\ln\frac{(s+b)(1 + \delta^2 b)}{\,b + \delta^2 b (s+b)\,} - \frac{1}{\delta^2} \ln\left(1 + \delta^2\frac{ s}{\,1 + \delta^2 b\,}\right) \Bigg]}\,. 
\label{eq:Zdisc_sys}
\end{equation}
This result is derived in Ref.~\cite{gorin} (their Eq.~(27)). It explicitly shows how the presence of a background uncertainty ($\delta > 0$) reduces the significance compared to the ideal case of no systematics. Indeed, in the limit of no systematic uncertainty ($\delta \to 0$), Eq.~(\ref{eq:Zdisc_sys}) reduces to the well-known formula for discovery significance without systematics:
\begin{equation}
\lim_{\delta \to 0} Z_{\text{disc}} = \sqrt{\,2\Big[ (s+b)\ln\left(1 + \frac{s}{b}\right) - s \Big]}\,,
\label{eq:Zdisc_nosys}
\end{equation}
which is the formula originally given in Ref.~\cite{cowan} for the median significance of a signal of size $s$ on top of background $b$ (for large $b$). Equation~(\ref{eq:Zdisc_nosys}) can be obtained from Eq.~(\ref{eq:Zdisc_sys}) by expanding for small $\delta$: the second term in Eq.~(\ref{eq:Zdisc_sys}) vanishes as $\delta \to 0$ and the first term simplifies using $\ln\frac{(s+b)(1+\delta^2 b)}{b + \delta^2 b (s+b)} \to \ln(1 + s/b)$.

\subsection{Exclusion Significance (Upper Limits): Analytical Formula}
For setting upper limits, we are interested in how confidently one can exclude a signal hypothesis (typically $\mu=1$) in the case that no true signal is present. In other words, we consider the background-only scenario ($\mu'=0$ as the true state of nature) and ask for the median significance with which we can reject $\mu=1$. This is sometimes called the sensitivity for exclusion or the power to exclude the signal hypothesis.

We focus on the test statistic $q_1$, which corresponds to Eq.~(\ref{eq:qmu_def2}) with $\mu=1$. Following a procedure analogous to the discovery case, one derives the expression for $q_1$ for given observed $n, m$ by inserting Eqs.~(\ref{eq:muhat})--(\ref{eq:bhathat}) into $-2\ln \lambda(1)$. The result (after simplification) for the profile likelihood ratio test of $\mu=1$ is~\cite{gorin}:
\begin{equation}
q_1 = 
\begin{cases}
-2 \Big[ n \ln \frac{s + \hat{\hat{b}}(1)}{n} + m \ln \frac{\tau \hat{\hat{b}}(1)}{m} - (s + \hat{\hat{b}}(1)) + n - \tau \hat{\hat{b}}(1) + m \Big]\,, & \hat{\mu} \le 1,\\[1ex]
0\,, & \hat{\mu} > 1~,
\end{cases}
\label{eq:q1_expr}
\end{equation}
where $\hat{\hat{b}}(1)$ is the conditional MLE of $b$ assuming $\mu=1$ (given explicitly in Eq.~(\ref{eq:bhathat}) for $\mu=1$). The condition $\hat{\mu} \le 1$ simplifies to $n - m/\tau \le s$ (using Eq.~\ref{eq:muhat}); if the observed excess in $n$ above the background estimate $m/\tau$ is larger than the expected signal $s$, then $\hat{\mu}>1$ and by definition $q_1=0$ (we would not exclude $\mu=1$ if data show an upwards fluctuation beyond the signal expectation).

To find the median exclusion significance, we evaluate $q_1$ on the Asimov dataset with $\mu'=0$ (background-only assumption for sensitivity). That means we set $n = b$ and $m = \tau b$. Under these conditions, $\hat{\mu}=0$ and $n - m/\tau = 0 \le s$ is satisfied (since $s$ is positive), so we use the first case of Eq.~(\ref{eq:q1_expr}). Plugging in the Asimov values and simplifying leads to an expression for the Asimov test statistic $q_{1,A}$. The median exclusion significance $Z_{\text{excl}}$ is then given by $\sqrt{q_{1,A}}$. After considerable algebra, one finds the following formula for $Z_{\text{excl}}$ including the effect of a background uncertainty~\cite{gorin}:\footnote{
While deriving Eq.~(\ref{eq:Zexcl_sys}) for the exclusion significance, we consulted auxiliary notes kindly shared by Prof.~Abdulkadir~Senol - the author of a recent phenomenological study~\cite{Senol}. In the published version of that work, Eq.~(9) contains a typographical bracketing error (the closing square bracket is misplaced). We discussed this point with Prof.~Abdulkadir~Senol in private correspondence; he acknowledged the oversight and indicated it would be corrected in an updated version. In the present paper we use the corrected expression consistent with our first-principles derivation. Our implementation and all numerical results in this paper therefore use this corrected form. We thank Prof.~Senol for sharing clarifying materials and for confirming the bracket placement issue.}

\begin{equation}
Z_{\text{excl}} = \sqrt{\,2 \Bigg[\, s - b \ln\frac{b + s + x}{2b} - \frac{1}{\delta^2} \ln\frac{b - s + x}{2b} \,\Bigg] - (b + s - x)\Bigg(1 + \frac{1}{\delta^2 b}\Bigg) }\,. 
\label{eq:Zexcl_sys}
\end{equation}
In this expression, $x$ is a function of $s$, $b$, and $\delta$ defined as 
\[
x = \frac{\sqrt{(b+s)^2 - 4\,\delta^2 b^2 s}}{\,1 + \delta^2 b\,}\,. 
\] 
Equation~(\ref{eq:Zexcl_sys}) corresponds to Eq.~(39) in Ref.~\cite{gorin}. While this formula is rather complex, it reduces to a much simpler form in the absence of systematics. Taking $\delta \to 0$ (which implies $x \to b+s$), the median exclusion significance becomes
\begin{equation}
\lim_{\delta \to 0} Z_{\text{excl}} = \sqrt{\,2\Big[\, s - b \ln\left(1 + \frac{s}{b}\right) \Big]}\,. 
\label{eq:Zexcl_nosys}
\end{equation}
This is the asymptotic formula for exclusion significance when the background is known exactly (cf. Ref.~\cite{cowan}). It differs from the discovery formula (\ref{eq:Zdisc_nosys}) in the placement of the logarithm: here $s - b\ln(1+s/b)$ appears, whereas in the discovery case we had $(s+b)\ln(1+s/b) - s$. In the limit $s \ll b$, both formulas (\ref{eq:Zdisc_nosys}) and (\ref{eq:Zexcl_nosys}) approximate the simpler result $Z \approx s/\sqrt{b}$ (as expected in the regime of small signal on large background), but for larger $s$ the two scenarios are distinct. 

The complete formulas (\ref{eq:Zdisc_sys}) and (\ref{eq:Zexcl_sys}) provide a way to estimate the sensitivity of an experiment including the degrading effect of systematic uncertainties in the background estimate. These are the formulas implemented in STC for quick calculations of expected significance and confidence limits.

\subsection{Generalized Significance Formulas}

The conclusions given in the previous section can be extended to the case of $N$ signal and $M$ control bins~\cite{Basso}. Here, bins are related to a histogram distribution of a observable collected from an experiment. It is assumed that the numbers of signal events $s_1, s_2, \ldots, s_N$ in each of the $N$ signal bins correspond to the same signal strength $\mu$, and also that the control region is orthogonal to the signal and independent of it. For $i=1,\ldots,N$ signal bins, there are matrices $\boldsymbol{\tau}^i\,;\, i=1,\ldots,N$, where $\boldsymbol{\tau}^i$ is a matrix matching the backgrounds of $i$-a $k$ signal bin with $M$ control bins (for example, $\tau_{jk}^i$ compares the $k$ background from the $i$ signal bin with the $j$ control bin). In this case, the likelihood function:

\begin{equation}
    \label{eqn:nsr_mcr_L}
    L(\mu,\boldsymbol{B}) = \prod_{i=1}^{N} \prob{n_i}{\mu s_i + \sum_{j=1}^{M}b_{ji}}\cdot \prod_{j=1}^{M} \prob{m_j}{\sum_{j^{\prime}=1}^{M} \tau_{jj^{\prime}}^{i^\prime} b_{j^{\prime}i^{\prime}}} \,,
\end{equation}
where $\prob{n}{\lambda}$ is the probability of a Poisson distribution, $\boldsymbol{B}$ is the matrix of backgrounds in signal bins ($b_{ji}\equiv[\boldsymbol{B}]_{ji}$ corresponds to the $j$th background in $i$th signal bin), and $i^{\prime}$ is any integer from 1 to $N$. For certainty, you can put $i^\prime = 1$.

Then, by calculating the maximum likelihood estimates of $\hat{\mu}$ and $\hat{\boldsymbol{B}}$ and using the Asimov dataset, we can obtain an expression for the expected significance of the discovery:

\begin{equation}
    \label{eqn:nsr_mcr_z0}
    Z_{disc} = \sqrt{-2\br{\sum_{i=1}^{N} \cbr{n_i\ln\!\br{\frac{\sum_{j=1}^{M}\bhh_{ji}}{n_i}} + n_i - \sum_{j=1}^{M}\bhh_{ji}} + \sum_{j=1}^{M} \cbr{m_j\ln\!\br{\frac{[\boldsymbol{\tau}^1\cdot\doublehat{\boldsymbol{B}}]_{j1}}{[\boldsymbol{\tau}^1\cdot\boldsymbol{B}]_{j1}}} + [\boldsymbol{\tau}^1\cdot(\boldsymbol{B} - \doublehat{\boldsymbol{B}})]_{j1} }} } \,,
\end{equation}
where $\doublehat{\boldsymbol{B}}$ is given by the expression:
\begin{equation}
    \label{eqn:nsr_mcr_bhh_eqns}
    \sum_{i=1}^{N}\frac{\tau^1_{1\ell}}{\tau^i_{1\ell}}\cdot\br{\frac{n_i}{\sum_{j=1}^{M}\bhh_{ji}} - 1} + \sum_{j=1}^{M} \tau^1_{j\ell} \cdot \br{\frac{m_j}{[\boldsymbol{\tau}^1\cdot\doublehat{\boldsymbol{B}}]_{j1}} - 1} = 0 \,;\, \ell=1,{\ldots},M\,.
\end{equation}

\section{Monte Carlo calculations}

Monte Carlo simulations are a key feature of the \textit{StatTestCalculator} (STC) tool, offering a precise and flexible approach to calculate statistical significance and upper limits in high-energy physics analyses. These simulations generate distributions of test statistics by performing repeated pseudo-experiments, providing exact calculations and accounting for the statistical uncertainty arising from the finite number of simulated trials.

\subsection{Overview of Monte Carlo Simulations in STC}

Monte Carlo simulations in STC involve the generation of synthetic data, known as \textit{toy experiments}, based on predefined signal and background distributions. These toy experiments simulate random fluctuations of the signal and background, incorporating both systematic and statistical uncertainties. For each generated dataset, STC computes a test statistic, such as \( q_0 \) or \( q_\mu \), and subsequently uses the distribution of these test statistics to determine p-values and significance levels.

\subsection{Monte Carlo Simulation Workflow}

The general workflow of a Monte Carlo simulation in STC can be broken down into the following steps:

\begin{itemize}
    \item \textbf{Toy Experiment Generation:} For each simulation (or \textit{toy}), STC generates synthetic data by drawing signal and background events from the appropriate distributions. The signal events are drawn from a Poisson distribution with a mean that is determined by the signal strength parameter \( \mu \) and the expected signal histogram. The background events are also drawn from a Poisson distribution based on the background histogram. If systematic uncertainties are included, these are applied to both signal and background distributions, either as shape-of-distribution uncertainty or a number of events in a single bin uncertainty.
    \item \textbf{Computation of Test Statistics:} Once the toy data is generated, the test statistic is calculated based on the likelihood function of the observed data. The test statistic could be \( q_0 \), which compares the likelihood of the background-only hypothesis against the signal-plus-background hypothesis, or \( q_\mu \), which is used for setting upper limits on the signal strength \( \mu \). These statistics are computed for each toy experiment.
    \item \textbf{Distribution of Test Statistics:} The test statistics computed from all the toys are collected into a distribution. This distribution represents the expected variation of the test statistic under the null hypothesis (background-only). It is then compared with the test statistic computed from the actual observed data.
    \item \textbf{Significance Calculation:} After the distribution of test statistics is calculated the p-value can be computed. The criterion for p-value calculation is the value of test statistic calculated from the observed data. The integral is calculated from the observed value of test statistic up to the end of the distribution.
\end{itemize}

\subsection{Handling Systematic Uncertainties}

STC allows for the inclusion of systematic uncertainties via either a lognormal or Gaussian distribution, which can be specified by the user.  $\tau$ parameter defined in previous section is used for the incorporation of systematics. Each time the toy data is generated, number of events is mulitplied by the $\tau$-factor which is drawn from the specified distribution of the systematic uncertainties (e.g. normal or lognormal). 
\
In case of the shape-of-distribution systematic uncertainty, number of toy events in each bin is multiplied by one scalar value. On the other hand, in case of the systematic uncertainty of the number of events in a single bin, the vector of values of parameter $\tau$ is drawn from the systematic distribution. The size of the vector equals to the number of bin, and each number of events in single bin is multiplied by its own $\tau$-factor.

\subsection{Upper Limit Calculation via Monte Carlo Simulations}

Monte Carlo simulations in STC are also used for calculating upper limits on the signal strength parameter \( \mu \). The process for calculating upper limits involves scanning through a grid of possible values for \( \mu \). For each value of \( \mu \), the following steps are performed:

\begin{enumerate}
 
    \item Toy experiments are generated for each value of \( \mu \).
    \item The test statistic \( q_\mu \) is computed for each toy experiment.
    \item The distribution of \( q_\mu \) is compared with the observed value of the test statistic, and the p-value is calculated for each value of \( \mu \).
    \item The smallest value of \( \mu \) for which the p-value falls below the chosen confidence level (e.g., 95\%) is taken as the upper limit on the signal strength \( \mu \).
\end{enumerate}

This method allows for an exact calculation of the upper limit, accounting for both statistical and systematic uncertainties, and provides more reliable results compared to asymptotic approximations.

\section{Implementation}
The StatTestCalculator is implemented in Python and structured to handle both asymptotic calculations and Monte Carlo simulations within a unified framework. The core of the tool revolves around evaluating the likelihood functions and test statistics described in the previous section for a given dataset or for pseudo-experiments. Key implementation details include:

\textbf{Asymptotic calculations:} STC directly uses the analytical formulas for $q_0$, $q_\mu$, $Z_{\text{disc}}$, and $Z_{\text{excl}}$. For instance, given user inputs of expected signal $s$, expected background $b$, and relative background uncertainty (which can be specified via $\delta$ or by providing a control region setup), STC computes $Z_{\text{disc}}$ from Eq.~(\ref{eq:Zdisc_sys}) and $Z_{\text{excl}}$ from Eq.~(\ref{eq:Zexcl_sys}). In the absence of systematics, the code automatically simplifies to Eqs.~(\ref{eq:Zdisc_nosys}) and (\ref{eq:Zexcl_nosys}). These formulas allow users to obtain instantaneous estimates of significance and exclusion power without needing to generate toy experiments.

\textbf{Monte Carlo simulations:} For more exact calculations or validation of the asymptotic approximations, STC can perform pseudo-experiments. The user can specify the number of toy experiments to generate. For each toy, the tool randomly generates data (e.g. drawing signal and background events from the set distribution Poisson($\mu s \ + \ \tau b$) if systematics are included or Poisson($\mu s + b$) if not and then computes the test statistic ($q_0$ or $q_\mu$) for that toy dataset. The distribution of $q_0$ or $q_\mu$ from many toys is then used to determine $p$-values and significances. The code ensures that the test statistic definitions follow the one-sided conventions: for example, any negative $\hat{\mu}$ results in $q_0$ being set to 0 in that toy (see Eqs.~(\ref{eq:q0_def}) and (\ref{eq:qmu_def2}), implemented as a \verb|max(0, ...)| operation in the code). The output of the Monte Carlo can provide an exact significance value (with statistical uncertainty from the finite number of toys).

\textbf{Systematic uncertainties:} STC allows two primary modes for incorporating systematics: (i) \emph{normal} (Gaussian) and (ii) \emph{lognormal} distributions for uncertainty in background and/or signal. In practice, these are implemented by treating $b$ (and potentially $s$ if signal systematics are considered) as random variables that modify the mean event counts. For example, a lognormal uncertainty on $b$ can be implemented by fluctuating $b$ by a lognormal factor in each toy experiment.  

\textbf{Computing a significance of discovery:} To calculate the exact value of a significance of discovery the code simulates a distribution of $q_0$ test statistic. Then it computes the p-value using the the value $q_{0, obs}$ calculated from the obtained data as a criterion as it shown in fig.1.
\begin{figure}
    \centering
    \includegraphics[scale=0.8]{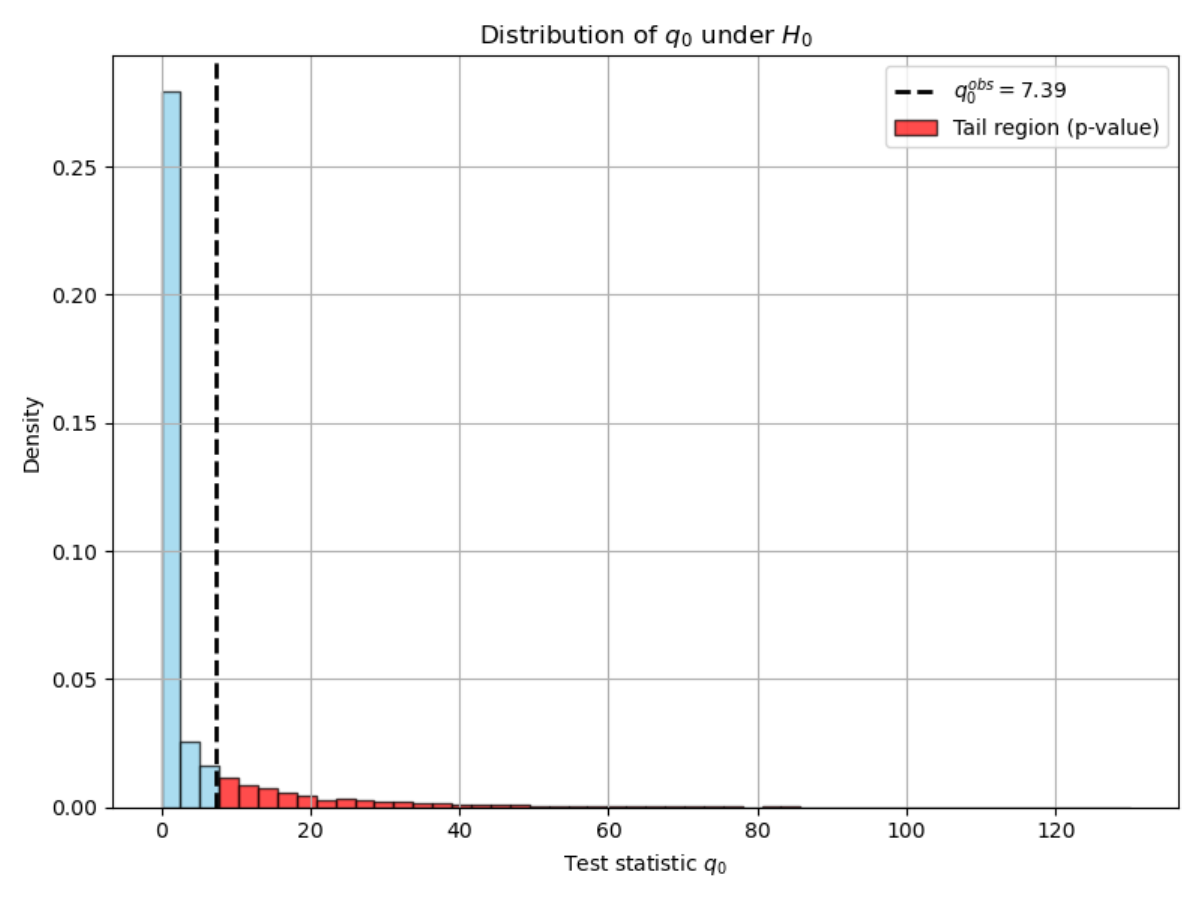}
    \caption{The example of Monte Carlo simulated distribution of test statistic $q_0$}
    \label{fig:q0}
\end{figure}

\textbf{Setting an upper limit:} To find the exact value of signal strength parameter $\mu$ the code scans through the grid of the possible values specified by the user. The iterating process calculates the $Z_\mu$ significance for each value from the grid until the set confidence level is reached as shown in fig.1:
\begin{figure}[h]
    \centering
    \includegraphics[scale=0.8]{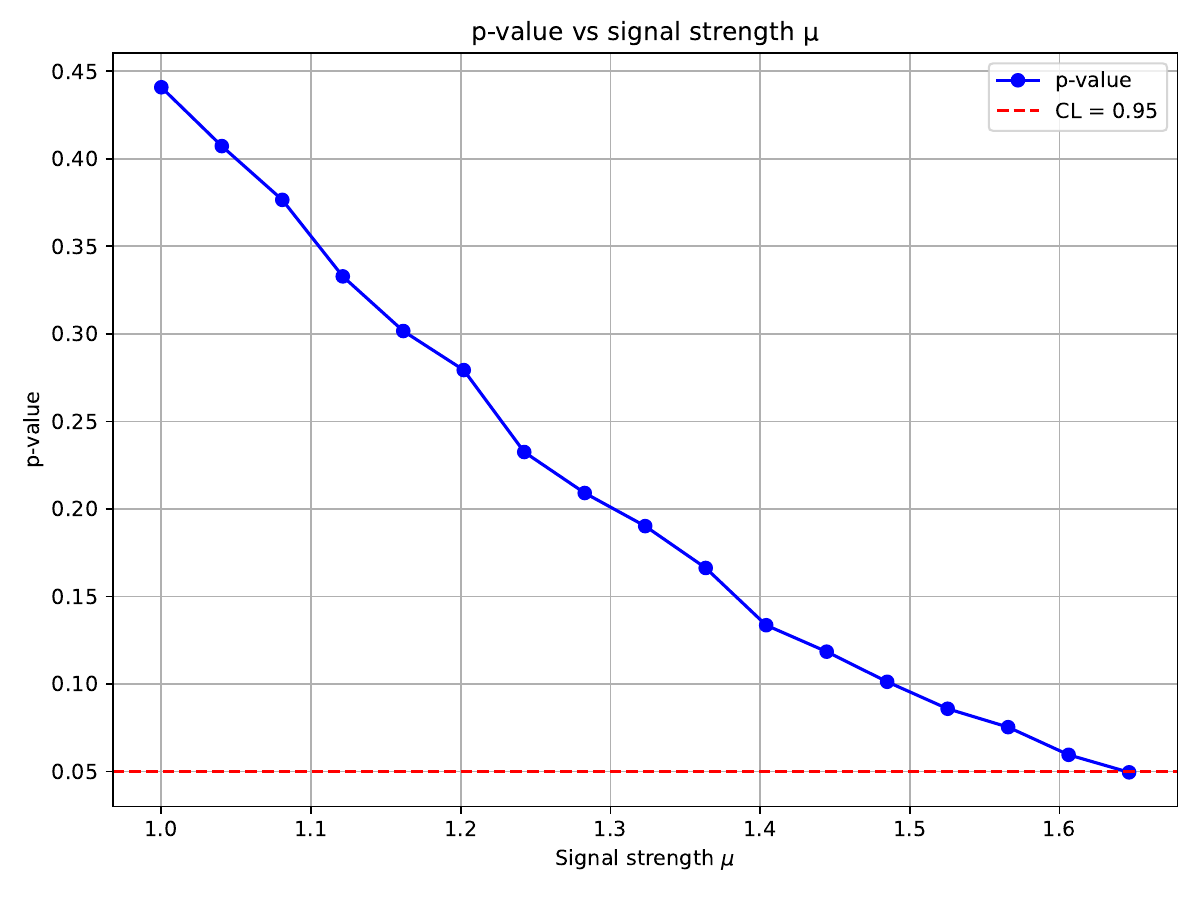}
    \begin{center}
        \caption{The convergence of $\mu$ parameter for each iteration of Monte Carlo simulations through the specified grid. The set confidence level of 95\% is reached for the value of $\mu \approx 1.7$.}
    \end{center}
    
    \label{fig:mu_conv}
\end{figure}

\textbf{User extensibility:} The code is written in a modular way, allowing advanced users to plug in custom definitions. For example, users can define their own test statistic if needed (beyond the provided $q_0$ and $q_\mu$) or even replace the likelihood with a different functional form (for instance, a shape analysis across multiple bins, which STC also supports as a generalization of the formulas~\cite{gorin}). STC’s design uses object-oriented principles: a class instance holds the data and model parameters, and methods like \verb|calculate_asymptotic_significance| or \verb|monte_carlo_hypotest| perform the desired calculation. This design makes it straightforward to add new methods or extend existing ones.

\textbf{Visualization and outputs:} In addition to numerical results, STC provides plotting functions. It can plot the input distributions (data vs. expected signal and background) for quick checks. It can also plot the distribution of the test statistic from toys. This is useful for verifying the theoretical asymptotic distribution. The results of comparisons (e.g. between STC and other tools, or between asymptotic and toy results) can be visualized directly, as we show in the next section.

Overall, the implementation emphasizes clarity and correctness, using well-tested libraries (such as NumPy for random sampling and SciPy for the normal CDF $\Phi$ and its inverse) to ensure numerical accuracy. The source code is available on GitHub for transparency and community contributions.

\section{Cross-checks and comparison}
We have validated the StatTestCalculator extensively on simplified scenarios, and we have compared its outputs with those of the widely-used CMS \textsc{Combine} tool~\cite{combine}. In this section, we present a few representative results demonstrating the performance and accuracy of STC.

\medskip
\noindent \textbf{Expected significance of a discovery:} We first examine the calculation of the expected significance of discovery. We consider a counting experiment with an expected background $b=100$ events and calculate the significance for different number of signal events $s=10, 20, 30, ..., 50$. We then compare the calculations obtained using STC asymptotic and MC calculations with the results from \textsc{Combine}. For each experiment we generate $10^5$ pseudo-experiments (toys) using STC. We also use different relative systematic uncertainty value for the valid comparison. The agreement between STC calculation and \textsc{Combine} is satisfying and shown in fig.3.

\begin{figure}[h]
\centering
\begin{subfigure}{0.45\textwidth}
    \centering
    \includegraphics[width=\textwidth]{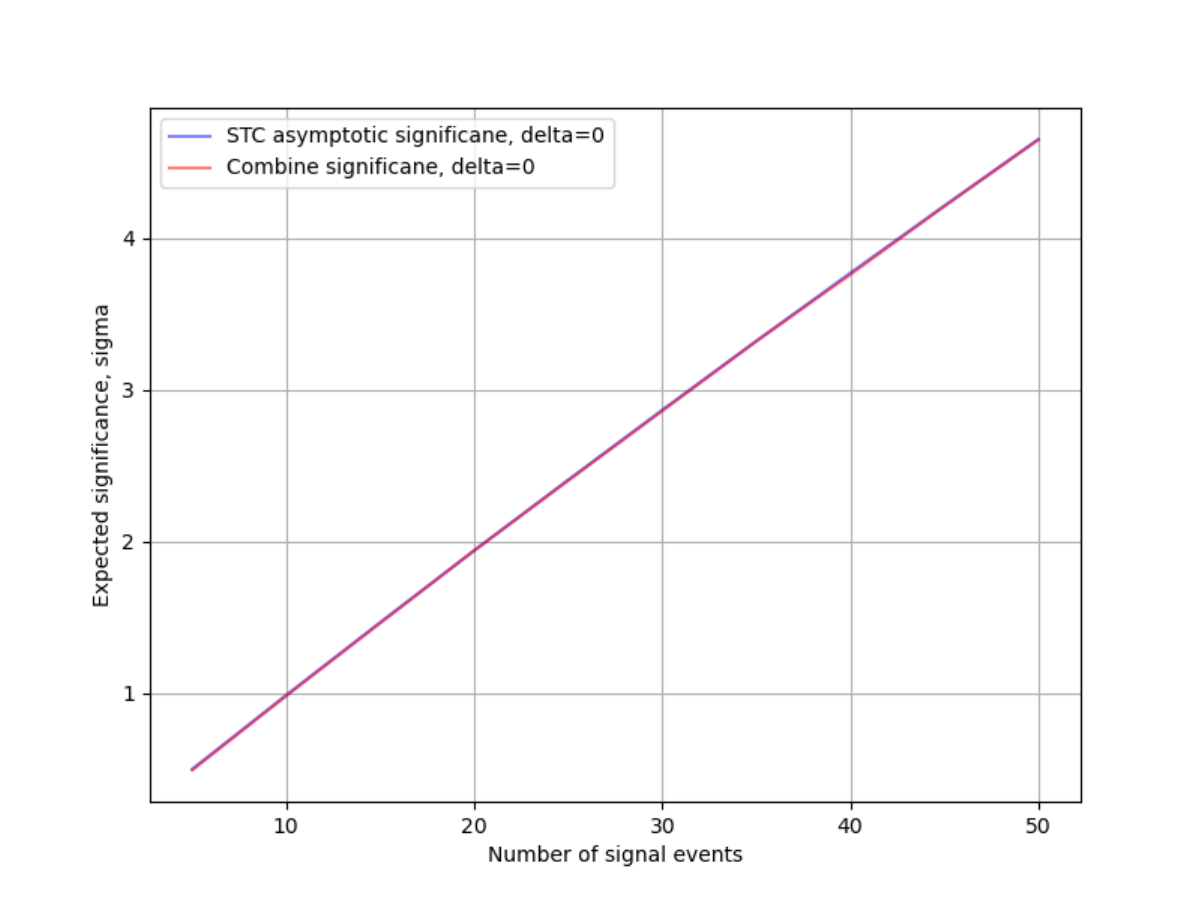} 
    \caption{Comparison of asymptotic calculations with no systematic uncertainty}
    \label{fig:subim1}
\end{subfigure}
\hspace{0.05\textwidth} 
\begin{subfigure}{0.45\textwidth}
    \centering
    \includegraphics[width=\textwidth]{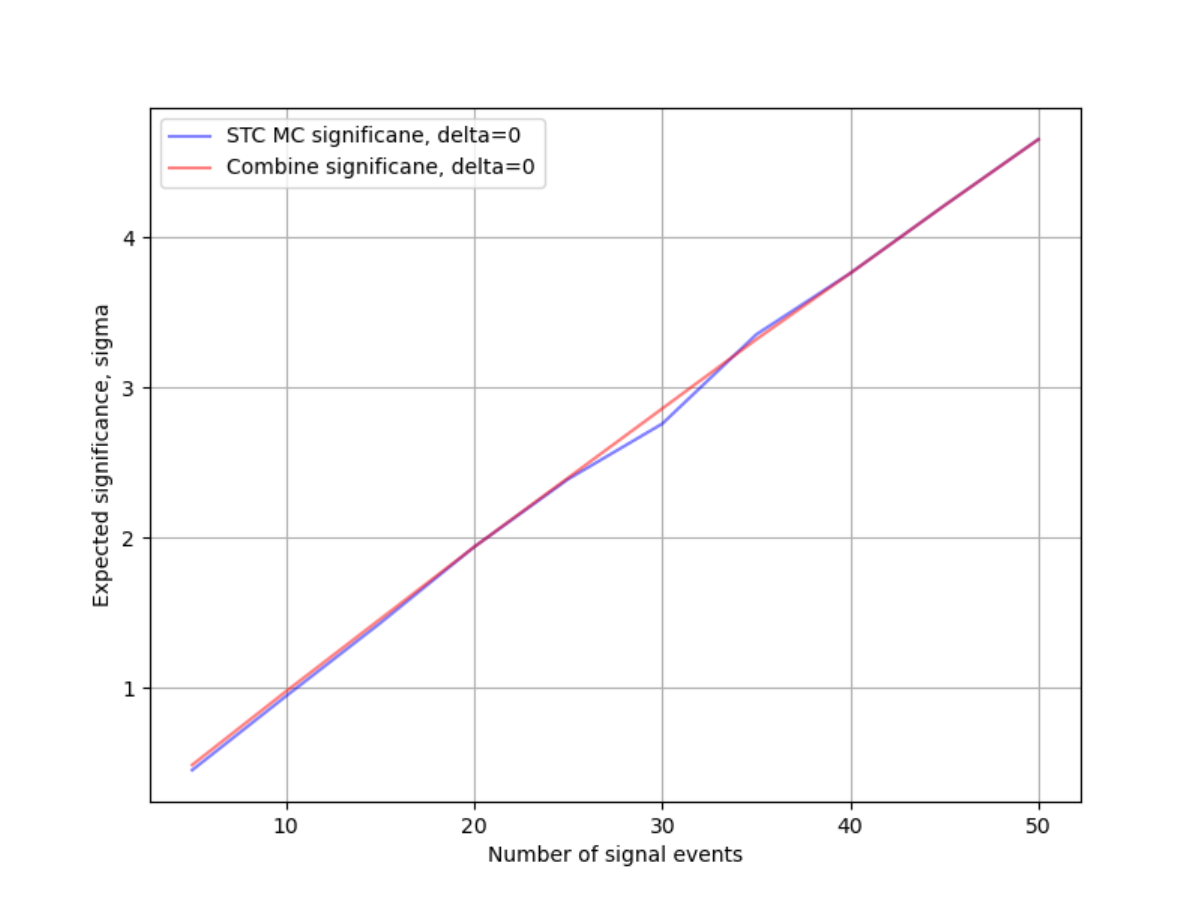}
    \caption{Comparison of Monte Carlo calculations with no systematic uncertainty}
    \label{fig:subim2}
\end{subfigure}

\vskip\baselineskip 

\begin{subfigure}{0.45\textwidth}
    \centering
    \includegraphics[width=\textwidth]{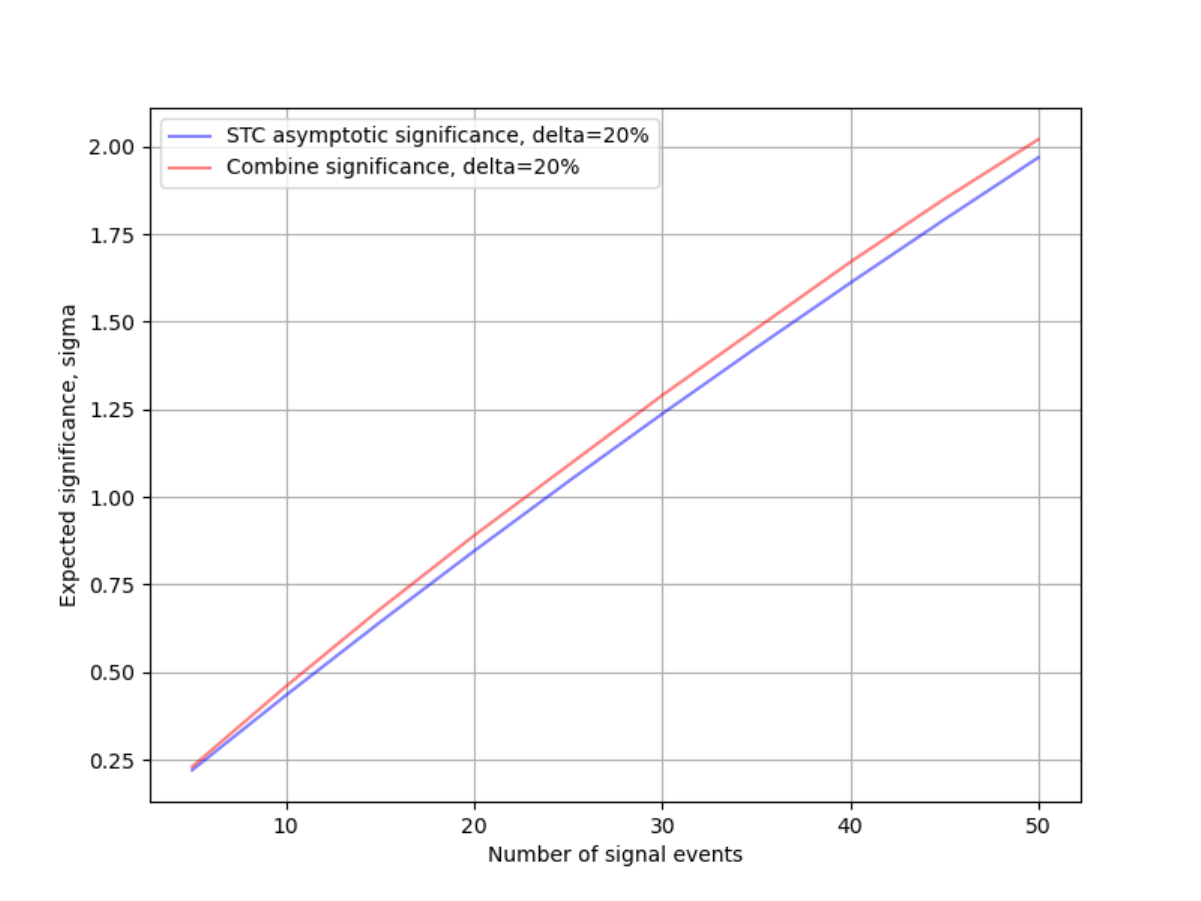}
    \caption{Comparison of asymptotic calculations with 20\% systematic uncertainty}
    \label{fig:subim3}
\end{subfigure}
\hspace{0.05\textwidth} 
\begin{subfigure}{0.45\textwidth}
    \centering
    \includegraphics[width=\textwidth]{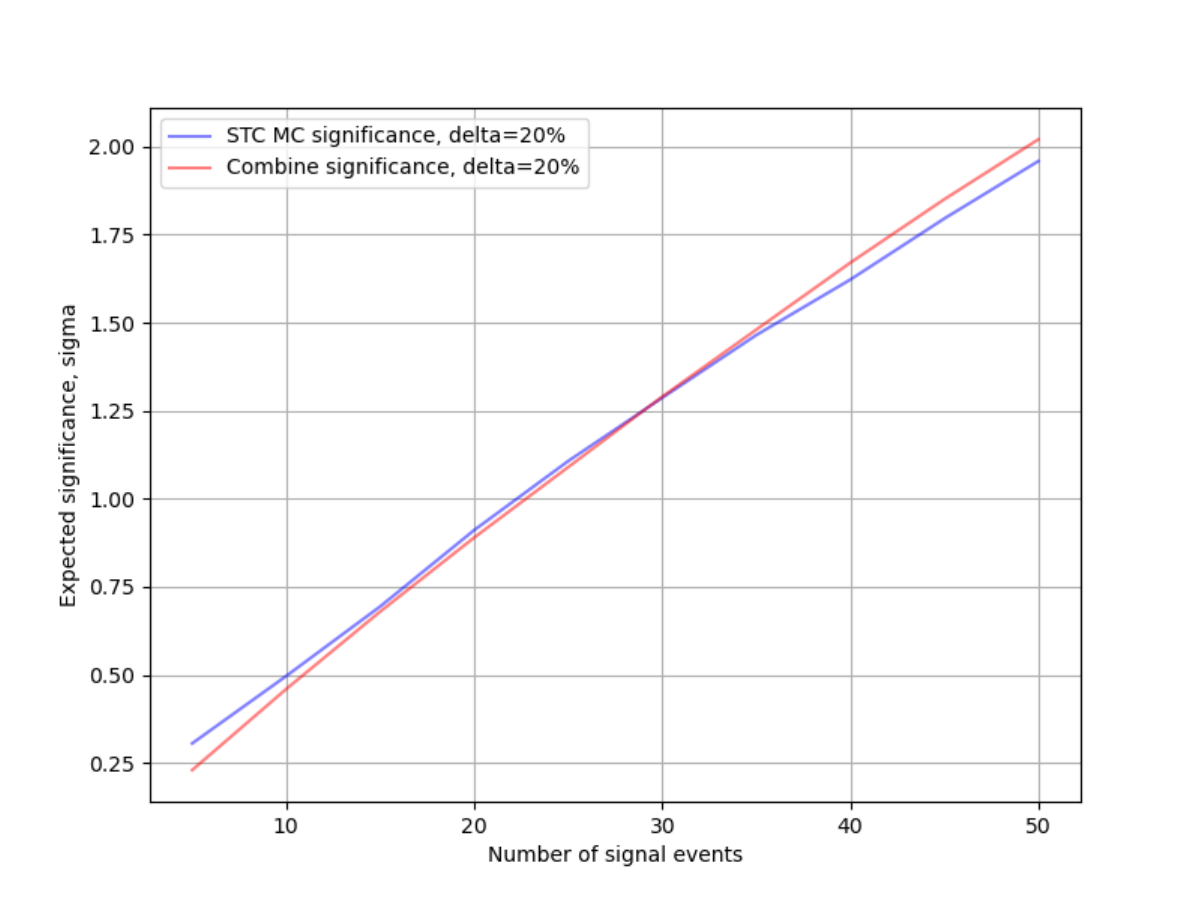}
    \caption{Comparison of asymptotic calculations with 20\% systematic uncertainty}
    \label{fig:subim4}
\end{subfigure}

\caption{Comparison of STC calculations of the significance of discovery made with asymptotic formulae (a, c) and using Monte Carlo simulations (b,d) with and without the presence of relative systematic uncertainty}
\label{fig:image2}
\end{figure}

\noindent \textbf{Expected upper limits on $\mu$ parameter:} We perform a similar comparison for the case of setting upper limits. We take the background $b=100$ and examine how the 95\% confidence level (CL) upper limit on the signal rate $\mu$ (expressed as an exclusion significance for $\mu=1$) behaves. We calculate the expected upper limits for different number of signal events $s = 10, 20,3,...,50$. Using STC, we compute the median exclusion significance $Z_{\text{excl}}$ for $\mu=1$ both asymptotically (Eq.~(\ref{eq:Zexcl_nosys})) and by toy simulation, under the assumption of $\mu'=0$ (no true signal). We compare these with the expected limit obtained from \textsc{Combine}. Once again, as it can be seen on fig. 4, the asymptotic formula and Combine agree very well over the range, and the toy-based results confirm the accuracy. This indicates that STC can reliably reproduce the standard calculations for upper limit sensitivity and the significance of discovery.

\begin{figure}[h]
\centering
\begin{subfigure}{0.45\textwidth}
    \centering
    \includegraphics[width=\textwidth]{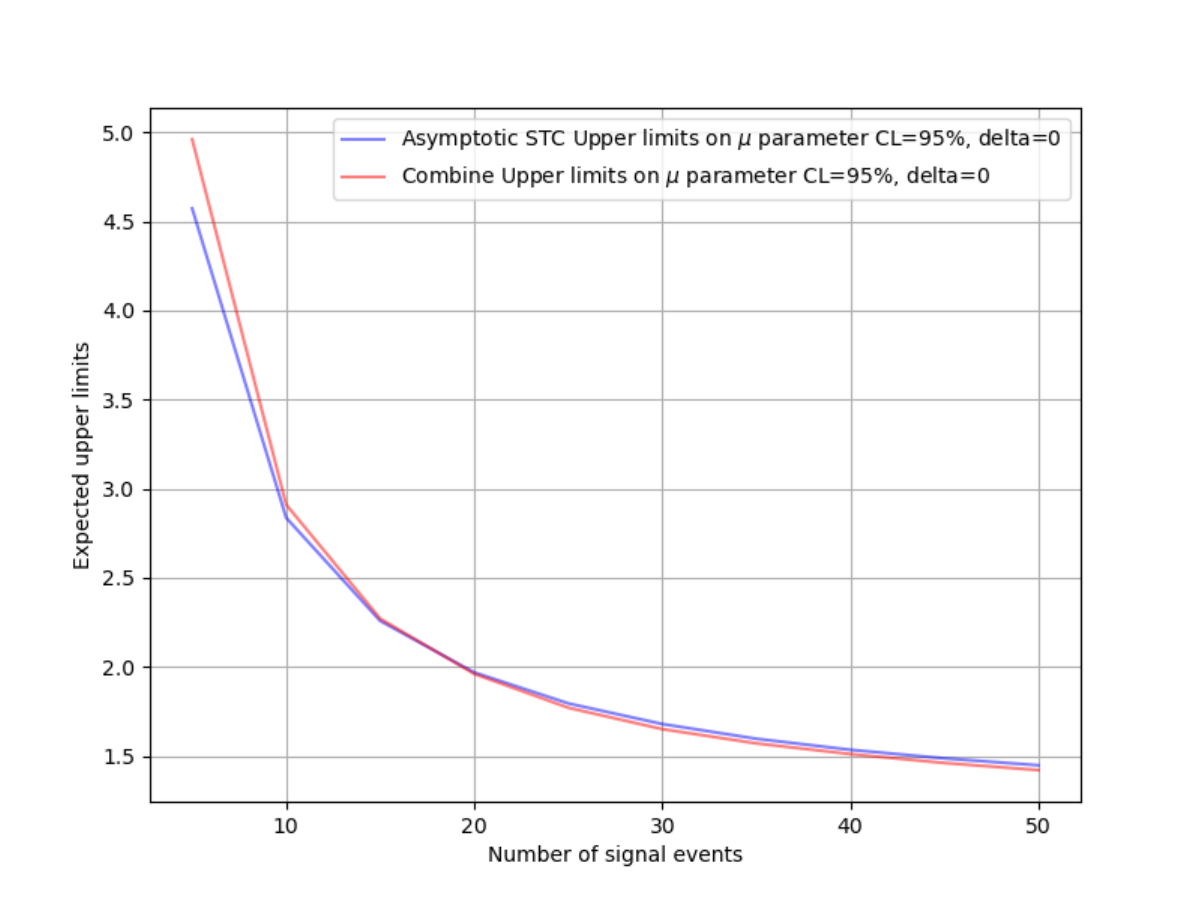} 
    \caption{Comparison of asymptotic calculations with no systematic uncertainty}
    \label{fig:subim1}
\end{subfigure}
\hspace{0.05\textwidth} 
\begin{subfigure}{0.45\textwidth}
    \centering
    \includegraphics[width=\textwidth]{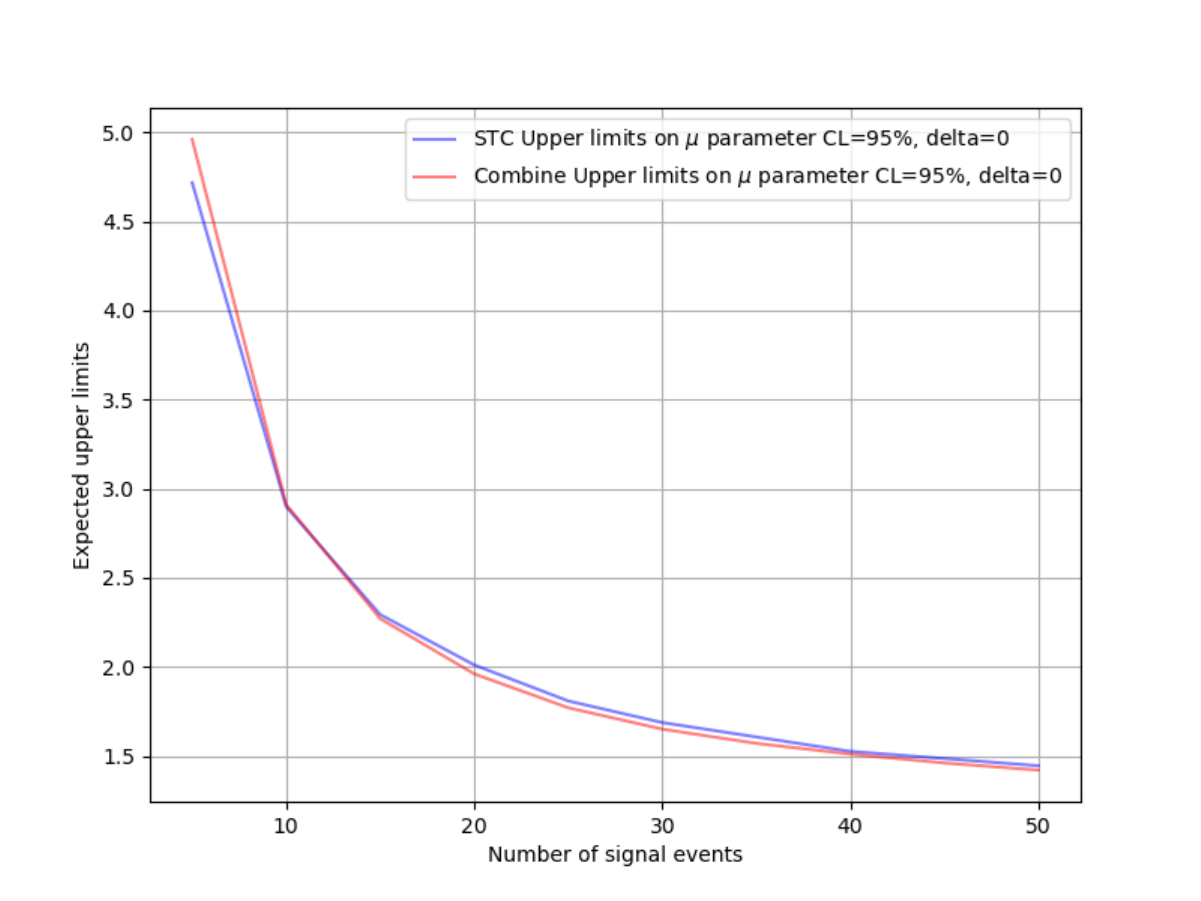}
    \caption{Comparison of Monte Carlo calculations with no systematic uncertainty}
    \label{fig:subim2}
\end{subfigure}

\vskip\baselineskip 

\begin{subfigure}{0.45\textwidth}
    \centering
    \includegraphics[width=\textwidth]{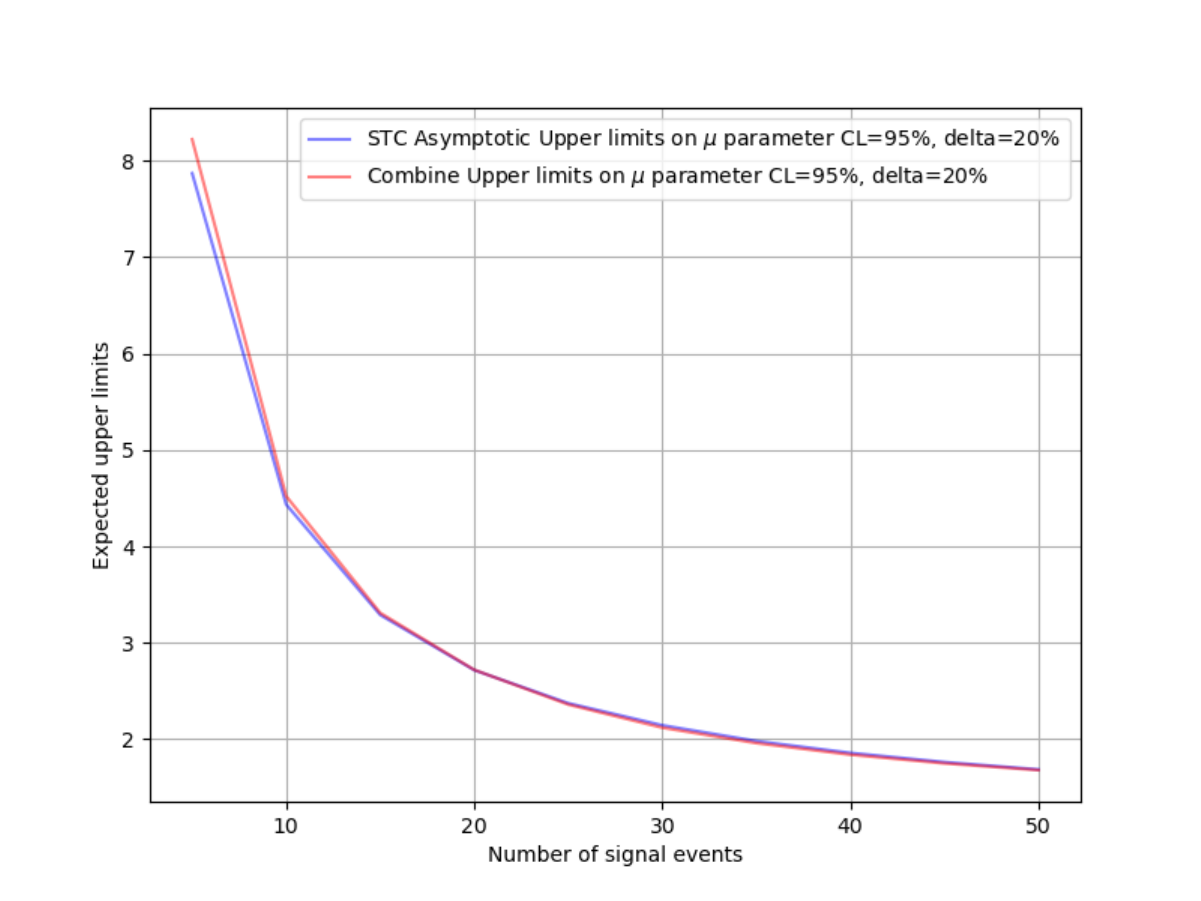}
    \caption{Comparison of asymptotic calculations with 20\% systematic uncertainty}
    \label{fig:subim3}
\end{subfigure}
\hspace{0.05\textwidth} 
\begin{subfigure}{0.45\textwidth}
    \centering
    \includegraphics[width=\textwidth]{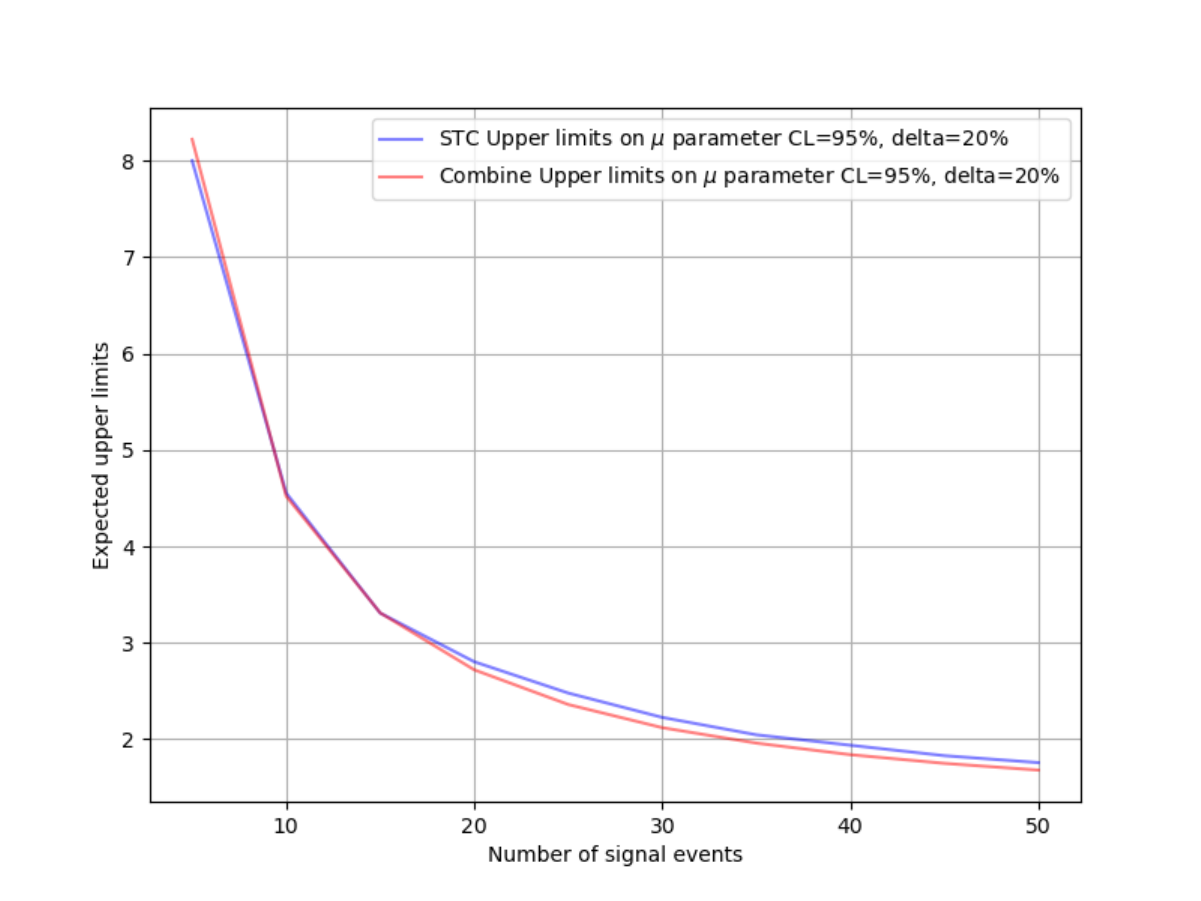}
    \caption{Comparison of Monte Carlo calculations with 20\% systematic uncertainty}
    \label{fig:subim4}
\end{subfigure}

\caption{Comparison of STC calculations of expected upper limits made with asymptotic formulae (a, c) and using Monte Carlo simulations (b,d) with and without the presence of relative systematic uncertainty}
\label{fig:image2}
\end{figure}

In addition to these specific tests, we have applied STC to more complex scenarios (such as multi-bin shape analyses with uncertainties) and compared with available results or closed-form expectations from Ref.~\cite{gorin}. STC was able to accurately compute significances in those cases as well. We also benchmarked the performance: STC’s asymptotic calculations run in negligible time (fractions of a second), and even Monte Carlo with $10^5$ toys completes in seconds to minutes for simple models, making it a practical tool for quickly exploring a wide parameter space of signal and background hypotheses.

The flexibility of STC is also noteworthy. For instance, one can easily switch the test statistic from the default profile likelihood $q_0$ to alternative definitions (a $CL_{s+b}$ or $CL_s$ construction, etc.) if needed for specialized studies. Similarly, one can incorporate a signal systematic uncertainty by providing an auxiliary dataset for signal or by analytically convolving a signal uncertainty distribution. These features make STC not only a validation tool but also a useful sandbox for developing new statistical methods.

\section{Conclusion}
We have developed a new statistical analysis tool, \textit{StatTestCalculator}, aimed at providing a general and user-friendly framework for significance estimation and limit setting in high energy physics. In this paper, we described the statistical methodology underpinning STC, including the formulation of likelihoods, profile likelihood ratio test statistics ($q_0$ for discovery and $q_\mu$ for limits), and the use of asymptotic theorems to obtain analytic formulas for significances. We also showed how these formulas can be extended to account for systematic uncertainties in the background, following the approach of \cite{cowan} and the recent derivations \cite{gorin}. The resulting expressions for the median discovery significance and exclusion significance were implemented in the tool for fast calculations.

The STC tool has a broad functionality: it can perform complete statistical analyses by calculating test statistics, determining $p$-values or CLs, and even producing plots for data and fit results. It supports quick exploratory estimates of sensitivity (using the asymptotic formulas) as well as more precise calculations via Monte Carlo. We demonstrated that STC’s outputs are in excellent agreement with those from the well-established \textsc{Combine} tool, for both discovery significance and upper limit determinations. This gives confidence that STC can serve as a reliable alternative for scenarios where a lightweight and easily customizable approach is desired.

The design of STC emphasizes versatility. Users can adapt it to their specific needs by plugging in custom models, including multiple channels or bins, adding uncertainties, or even defining new test statistics. This versatility means the tool can be applied not only to simple counting experiments but also to more complex analyses, such as those involving shape fits to histograms (an extension for which we have included formulae following Ref.~\cite{gorin}). We envision that STC will be helpful for physicists conducting preliminary sensitivity studies, cross-checks of official results, or educational purposes when learning about statistical tests in HEP.

\smallskip
The StatTestCalculator software, along with documentation and examples, is freely available as an open-source project on GitHub at \url{https://github.com/skottver/stattestcalculator}. We welcome the community to use the tool and contribute to its further development. 
\section*{FUNDING}
This study was conducted within the scientific program of the Russian National Center for Physics and Mathematics, section 5 «Particle Physics and Cosmology».

\section*{CONFLICT OF INTEREST}
The authors of this work declare that they have no conflicts of interest.



\end{document}